\documentclass[12pt,preprint]{aastex}
\slugcomment{Accepted for publication 
in {\it the Astrophysical Journal Letters}  and scheduled for v. 614, 2004, October 20 issue} 
\def\lax {\ifmmode{_<\atop^{\sim}}\else{${_<\atop^{\sim}}$}\fi} 
\def\gax {\ifmmode{_>\atop^{\sim}}\else{${_>\atop^{\sim}}$}\fi} 
\def\gtorder{\mathrel{\raise.3ex\hbox{$>$}\mkern-14mu
             \lower0.6ex\hbox{$\sim$}}}
\begin{document}

\title{Is M82 X-1 Really An Intermediate-Mass Black Hole? 
X-ray Spectral and Timing  Evidence}

\author{Ralph  Fiorito\altaffilmark{1} and Lev Titarchuk\altaffilmark{2}}

\altaffiltext{1}{University of Maryland, College
Park and NASA/GSFC, Greenbelt MD 20771; rfiorito@milkyway.gsfc.nasa.gov; 
rfiorito@UMD.edu}
\altaffiltext{2}{George Mason University/CEOSR,
Fairfax, VA 22030 and US Naval Research
Laboratory, Code 7655, Washington, DC 20375; ltitarchuk@ssd5.nrl.navy.mil and 
lev@lheapop.gsfc.nasa.gov}

\begin{abstract}

Ultra-luminous X-ray sources (ULXs) with apparent luminosities
up to 100's of times  the Eddington luminosity for a neutron star have been discovered
in external galaxies. The existence of intermediate mass black holes has been
proposed to explain these sources. We present evidence for an
intermediate-mass black hole in the ULX M82 X-1 based on the spectral features and
timing (QPO) properties of the X-radiation from this source. We revisited XMM
Newton and RXTE data for M82 X-1 obtained in 2001 and 1997 for XMM and RXTE
respectively. We show for these observations that the source is either in
transition or in a high/soft state with  photon spectral indices 2.1 and 2.7
respectively. We confirm the early determination of the QPO frequency
$\nu\approx 55$ mHz  in this source by Strohmayer \& Mushotzky  and identify this as
the low frequency QPO for the source. We apply a new method to determine the BH
mass of M82 X-1. The method uses the index-QPO low frequency correlation that
has been recently been established in galactic black hole candidates GRS
1915+105, XTE J1550-564, 4U 1630-47 and others. Using scaling arguments and
the correlation derived from consideration of galactic BHs, we conclude that
M82 X-1 is an intermediate BH with a mass of the order of 1000 solar masses.

\end{abstract}

\section{Introduction}
A number of external galaxies, notably ones with active star formation
regions, have revealed the presence of so called ultra-luminous X-ray sources
(ULXs) which one can operationally define as an X-ray source, which is not
coincident with the nucleus of its host galaxy and which has an apparent
luminosity in excess of an order of magnitude above the Eddington luminosity
for a neutron star ($L_{X}\gax 10^{39}$ erg~s$^{-1}$). The high luminosity of ULXs
have led to speculation that these objects may be intermediate ($10^{2}-10^{4})M_{\odot}$ BHs 
[e.g. Colbert \& Mushotzky 1999; Strohmayer \& Mushotzky 2003, hereafter SM03; Shrader \&
Titarchuk 2003, hereafter ST03; Miller et al. 2003] or, more conservatively, beamed sources of lower mass
(King et al. 2001). However, due to their distances the
prospects for dynamical measurement of the mass of ULXs are poor and other
means must be sought to identify their masses.

 ST03
present 
the analysis of observational data for
representative objects of several classes: GBHs, narrow-line Seyfert galaxies
(NLS1s) and ULXs. They apply a methodology, which uses the spectral
characteristics of the X-ray source radiation, to determine the mass. 
 Using a color
temperature from the bulk motion Comptonization  (BMC) model  
ST03 calculate the mass of
several GBHs. Their results are in agreement with the masses obtained by
optical or other methods when available. 
ST03 also
calculate masses in ULXs for seven sources. The most compelling cases for the
presence of intermediate-mass objects are  NGC 1313 and NGC 5408 objects
for which $\log(M/M_{\odot})=2.2,~3.0$ respectively.

There is  suggestions in the literature (see  King et al. 2001) that the radiation
from some of ULXs sources is beamed and thus the real luminosity is much smaller
than the Eddington luminosity for a solar mass BH (or neutron star). 
In fact, King et al. do not exclude the possibility that  {\it individual} ULX's may contain 
extremely massive BH. But they emphasize that the formation of large number of ultramassive BHs is problematic in
terms of evolutionary scenario of  binary systems.
The recent discovery of the fast variability [quasiperiodic oscillation (QPO)] of the X-ray radiation in
ULX M82 by SM03,  suggests that  
the X-ray emission area is quite compact in this particular source.
In fact, for a jet of a typical size about
$10^{13-14}$ cm one cannot expect the QPO frequency $\nu\simeq50$ mHz that is
observed by SM03 in M82 X-1. Thus, if the radiation from these variable sources
is isotropic, then the high inferred luminosity of M82-X1 requires an
intermediate-mass BH, i.e. a BH more massive than one that can be formed in
the collapse of a single normal star. 

If ULXs are indeed accreting BHs with properties presumably similar to GBHs,
similar observable properties, in particular, correlations of the X-ray spectral
indices and QPO frequencies in ULXs should be seen [see Vignarca, et.al.
(2003); Titarchuk and Fiorito (2004),
hereafter TF04; Titarchuk, Lapidus \& Muslimov (1998)].
TF04 
show how the mass of one GBH source can be used to determine
that of another, e.g. GRS 1915+105 and XTE J1550-564 which exhibit remarkably
similar QPO frequency-index correlation curves (shown here as Fig. 1). 
Once the mass of one object (e.g.
GRS 1915+105) is determined by a fit of theory to the measured QPO frequency-index
correlation curve, the mass of the other (XTE J1550-564) can be  found
by simply scaling, i.e. sliding the correlation curve for GRS 1915+105 along
the frequency axis until it coincides with the correlation curved of XTE
J1550-564, in this case by the factor 12/10. Thus the inferred mass of XTE
J1550-564 is 10/12 times less than that of GRS 1915+105. The mass
determination using the QPO frequency-index correlation fit is consistent with
 X-ray spectroscopic and dynamical mass determinations for these sources (see references in ST03).
Since QPO frequencies are inversely proportional to the mass of the
central object one can, in principle, determine the mass of a ULX using the
same kind of scaling as described above for GBHs. 
It is important to note that the BH mass determination method using the QPO-index correlation
is independent of the orientation of source.
In fact, the photon index of the Comptonization spectrum is a function of the Comptonization parameter only which
 is  a product of the  mean energy change at any   photon scattering and the mean number of
scatterings in the Compton cloud (corona) (see e.g. Sunyaev \& Titarchuk 1980).  

In \S 2 we present the results of our spectral and timing analysis of archival XMM and
RXTE data from the ULX source M82 X-1. In \S 3 we discuss our results of the
data, compare it to that presented by SM03 and employ our results to estimate
the mass of M82 X-1. Conclusions also follow in \S 3.

\section{DATA  ANALYSIS  and RESULTS}

We have revisited archival spectral and timing XMM and RXTE data from the ULX
source M82 X-1, which has been previously investigated by SM03  and Rephaeli and Gruber (2002)
 hereafter RG02.
We analysed four archival observations: (1) the XMM EPIC PN and MOS data from
the 30 ks observation May 5, 2001 (OBSIDS 0112290201) and (2) RXTE PCA data
from 1997 archival observations of RG03 \{OBSIDS: 20303-02-02(2.5 ks),
20303-02-03(4.76ks) and 20303-02-04(2.5 ks)\}. The latter two observations
show the presence of QPOs and a power law (PL) component in their energy
spectra, which we have reanalyzed in terms of the transition layer model
previously applied by us to the study of galactic black hole sources (TF04).
We were not able to identify the QPO reported by SM03 for OBSID
20303-02-02(2.5 ks).

\subsection{Energy spectra}
{\it XMM data.--} We extracted the spectrum of M82 X-1 from both the EPIC MOS and PN images of M82
in a circle of 18$^{\prime\prime}$  in radius  around the bright source (M82 X-1), using the latest version of the SAS data reduction software
and response matrices. We focused our attention on the energy range 3.3
keV~-~10 keV where interference from the soft diffuse component surrounding
the bright source is minimized. We find that this energy range gives more
consistent results for the spectral modeling and timing analysis than that
used by SM03 (2-10 keV), i.e. increasing the lower energy value from 2 to 3.3 keV produced 
the lowest value of resuduals in our energy spectral fits ($\chi_{red}^{2}\leq2$), and had virtually no effect on 
the $\chi_{red}^{2}$ and parameter values of our fits to the power spectrum when compared those using the energy
range 2-10 keV.
Using XSPEC (version 11.3), we find that an absorbed
power law model with a Gaussian iron line component and the Bulk
Comptonization model (BMC) (Titarchuk et al. 1997) with a Gaussian iron line
component fixed at 6.5 keV and a line width of 0.4 keV, give equally good fits
($\chi_{red}^{2}\leq1$), when the column density is left as a
free parameter. Since the real column density is unknown, because of the
obscuration effect of the surrounding diffuse emission surrounding M82 X-1, we
allow this parameter to be free. 
The $\chi_{red}^{2}$, and fitted
parameters showed little change with temperature up until  $kT=0.5$  keV. 
In order to obtain the
index, normalization and column density, 
we therefore fixed the value of kT
below this value, i.e. $kT=0.1$ keV. 
We found that for the best fit spectrum 
 the contribution of the Comptonized  component in the BMC spectrum is dominant. 
The BMC model spectrum  is a sum of
the (disk) black-body component and Comptonized black-body component, 
where $A/(1+A)$ is a relative weight of the Comptonized component. 
Because all values of parameter $A\gg 1$ are consistent with observations  we fixed $\log A$ to 5.
With these constraints we find that the power law index and
column densities from the PL and BMC fits give close to the same results for
all the observations investigated by us. We also also noticed that an absorbed
multi-temperature thermal disk model (diskBB) with a Gaussian, provides a
statistically good fit but that a high disk temperature $(T_{i}\sim3$ keV) is
required in agreement with the results of SM03. An example of a fit to our
reduced PN data using the BMC-Gaussian Fe Line model is given in Figure 2. A
comparison of fits to our reduced data for the PL and BMC with that produced
by SM03 for the energy band (3.3-10 keV) produces nearly identical results.
Fits to the MOS data [similar data is mentioned but not presented in (SM03)]
gives close to the same fitted parameters as the PN data fits.

\smallskip
{\it RXTE data.--}Spectra from M82 was also obtained from the archival observations of RG02
(using PCA Standard Products and the Epoch 3 response matrix).
A comparison of the free spectral fit parameters for the XMM EPIC PN, MOS and
RXTE PCA data is given in Table 1. Based on the photon indices obtained from
the XMM PN observation we conclude that the source was in a transition from
hard to soft state , i.e. characterized by an index between 1.5 and 2.5, for
the time period of the XMM observation and that for the period of the RXTE
data reported here the source was in a soft state (i.e spectral index $\sim2.5$).

\subsection{Timing Analysis}

{\it XMM data.}--We extracted a light curve from the EPIC\ PN instrument selected, from the same area A described in 
\S 2.1
in the energy band 3.3-10 keV with a 1 sec time bin and one 30 ks window which includes all the PN data. 
From this light curve we computed the power
spectra using the current release of the XRONOS program. The power spectrum
was rebinned a factor of 256 to obtain the results shown in Figure 3 (left panel) which
exhibits a clear QPO.
The high error in the red noise portion of the power spectra below 30mHz does
not allow a determination of the expected break frequency below the QPO frequency. Thus we
can not reliably identify the break frequency which is usually found with the
QPO low frequency in GBHs source from the XMM data (see Wijnands \& van der
Klis 1999). The data suggests but does not confirm the presence of a peak at approximately
100mHz, but the statistics are too poor to positively identify this frequency.
(see also the XMM data of SM03). Because of the severity of the red noise below 30 mHz we have, following SM03 chosen
to fit the power spectrum with a model consisting of a constant to account for the Poisson noise, a simple power law
to account for the red noise and  a Lorentzian to account for the presence of the QPO. Our fit to the data using this
model gives  a good fit $\chi_{red}^{2}=47$ for 58 dof, $\nu_{QPO}=58.5\pm 1.7$ mHz, $\Delta \nu_{QPO}/2=5.8\pm 1.5$ mHz and 
$A_{QPO}=0.011\pm 0.003$ which are in good agreement with the more accurate results obtained by SM03 using all three EPIC
instruments. Here $A_{QPO}$ represents the total rms power in the QPO,
and $\Delta\nu_{QPO}$ is the FWHM of the QPO. We checked the significance of the QPO by observing that $\chi_{red}^{2}=70$ when the QPO was excluded from the
model. For this value of $\chi_{red}^{2}=70$ using the F test  one  finds that   a probability $\approx 3\times10^{-5}$ 
for a random occurance of the QPO feature. These
results give us great confidence that the QPO is observed by the PN instrument (cf. SMO3).

{\it RXTE data.--}We extracted a light curve using good time intervals and 3-10 keV energy
selected power density spectra from the three RXTE OBSIDs listed above,
employing a 128 Hz sampling rate and only the top xenon layers of the
operating PCU's for each observation time interval. The IDL program RADPS,
written by C. Markwardt was used to produce the power density spectrum (PDS) averaged over several 1024 second
intervals for each light curve. 
We clearly identified single QPO's in the PDS
for two of the three OBSIDs studied by SM03. 
We fit
our PDS data to a sum of two Lorentzian peaks, the lowest or zero order
Lorentzian was picked to identify the red noise break frequency of the PDS. A
clearly identifiable break frequency at $26\pm 2.5$ mHz with $A_{br}=4.8\pm 0.4$ confirmed by a plot of
frequency time power vs frequency, coincident with a QPO at $106\pm 2$ mHz and $A_{QPO}=0.095\pm 0.018$ was
observed for OBSID 20303-02-04. A QPO at  $48.9\pm 1$ mHz and $A_{QPO}=0.04\pm 0.013$ was also observed for OBSID 20303-02-03,  
but a clearly  definable  break frequency was not observed for this OBSID.
Lorentzian line fits to the PDS for the former case is presented in Figure 3 (right panel).
We also looked for variation of the total variability as a function of energy
by observing the variability in two energy bands: 3-6 keV and 6-10 keV, but no
variation was observed.


\section{Discussion and Conclusions}
The spectral data for M82 X-1 definitely show the spectral features of
high/soft spectral phase when the photon indices $\Gamma$ are in the range of
$2-2.7$. For XMM PN OBSID and RXTE 20303-02-04 OBSID we find $\Gamma
=2.07\pm0.07$ and $\Gamma=2.67\pm0.1$ respectively; the former value is
characteristic of transition to the high/soft state, and the later to the
high/soft state. In these observations two QPO frequencies $\sim50$ mHz and
$\sim100$ mHz have been identified. 
We interpret and identify the observed $\nu_{QPO}\sim50$
mHz as the fundamental low frequency and $\nu_{2QPO}\sim100$ mHz as its first
harmonic. We note that the presence of one predominant QPO, i.e. 50 or 100 mHz
in different observations is not an unusual occurrence (see e.g. TF04) and can
be explained as the result of the local driving frequency conditions in the
coronal region; i.e. a resonance condition is established for one
particular eigenmode of the compact coronal region so that this mode is
predominantly observed. In other cases, for example  in GRS 1915+105 
(Fiorito et al. 2003)   the fundamental and first harmonics of
the low frequency QPO (along with break frequencies) are simultaneously observed. Such features are seen
in a variety of other GBH sources as well. Also the proximity of the inferred fundamental $\nu_{QPO}\sim50$ mHz to the observed break frequency,
$\nu_b\sim26$ mHz is similar to what is observed in GBH's. 

If we identify $\nu_{QPO}\sim 50$ mHz as the low frequency QPO frequency $\nu_{low}$ keeping in
mind that $\Gamma\sim2.7$ and $\nu_{low}\sim5$ Hz for 10 solar masses (see
Fig. 1, for the index-QPO frequency correlation in XTE 1550-564) and the fact
that $\nu_{low}$ is inversely proportional to $M$, we calculate $M\sim
5~(\mathrm{Hz})/0.050~(\mathrm{Hz})\times10=10^{3}$ solar masses. As we show
above (see also TF04) this scaling have been observed for galactic black holes
but this is the first time it has been applied to a ULX source (M82 X-1) to
estimate the BH mass.
This value of BH mass is consistent with  mass evaluations obtained using
the absolute normalization and color temperature of other ULXs (NGC 253, NGC
1399 X-2, X-4 and IC 342 X-1) 
 which have been analyzed by ST03.
We note the possibility that the 50 mHz QPO may not be the lowest intrinsic QPO frequency and that a lower one, which may be
obscured by the red noise at frequencies $\nu_{QPO}< 50$ mHz may exist. 
Therefore the frequency $\nu_{QPO}\approx 50$ should be interpreted 
as an upper limit and therefore the inferred mass as a lower limit.

To conclude, we have presented a reanalysis and new interpretation of XMM Newton and RXTE
data obtained from M82 X-1. This analysis presents via the BMC model spectral
photon indices 2.1-2.7 which are seen in the high/soft states of extragalactic
and galactic black holes and identification of $\nu_{QPO}\approx 50$ mHz as
an observable upper limit on the  low frequency QPO for M82-X1. Using this value for the  low QPO frequency,
the predominantly observed spectral index $\sim 2.7$ and the index-QPO
frequency correlation recently obtained for GBHs establishes  a lower limit the
mass of M82 X-1 of the order 1000 solar masses.
The demonstrated application of our method  which uses the low QPO frequency and the index of the power-law component of the
spectra  presents a new,
potentially powerful tool for determing the nature and mass of  ULXs.
However, the confirmation of a ULX as an intermediate BH awaits either the
simultaneous application of the two independent methods we have described
above, i.e. the ST03 method and  QPO low frequency-index correlation or
direct dynamical evidence.

R.F. gratefully acknowledges helpful discussions with T. Strohmayer, C.
Markwardt and L. Angelini.
L.T. acknowledges the support of this
work by the Center for Earth Observing and Space Research of George Mason University.



\begin{deluxetable}{rrrrrrrrrr} 
\tablecolumns{7} 
\tablewidth{0pc} 
\tablecaption{The best-fit spectral parameters.}
\tablehead{ 
\colhead{OBSID}&\colhead{Instrument}
  & \colhead{$n_{H(PL)}\times10^{22}$}  & \colhead{$n_{H(BMC)}\times10^{22}$}  & \colhead{$\Gamma_{PL}$}  & \colhead{$\Gamma_{BMC}$}  
& \colhead{$N_{BMC}\times(10^{-3})$} }
\startdata 
011201 & XMM-PN & $7.50\pm0.69$ & $7.50\pm0.69$ & $2.07\pm0.07$ & $2.07\pm0.07$ & $0.26\pm0.08$\\
011201 & XMM-MOS & $6.98\pm1.17$ & $7.42\pm1.21$ & $1.71\pm0.14$ & $1.82\pm0.16$ & $0.21\pm0.13$\\
20303-02-02 & XTE-PCA & $3.73\pm0.76$ & $4.43\pm1.53$ & $2.62\pm0.09$ &
$2.65\pm0.13$ & $1.13\pm0.67$\\
20303-02-03 & XTE-PCA & $4.09\pm0.78$ & $5.26\pm1.93$ & $2.46\pm0.09$ &
$2.52\pm0.13$ & $1.11\pm0.67$\\
20303-02-04 & XTE-PCA & $6.06\pm0.81$ & $5.21\pm0.85$ & $2.67\pm0.09$ &
$2.63\pm0.1$ & $1.35\pm0.55$
\enddata 
\end{deluxetable}


\newpage
\begin{figure}[ptbptbptb]
\includegraphics[width=5.1in,height=4.5in,angle=0]{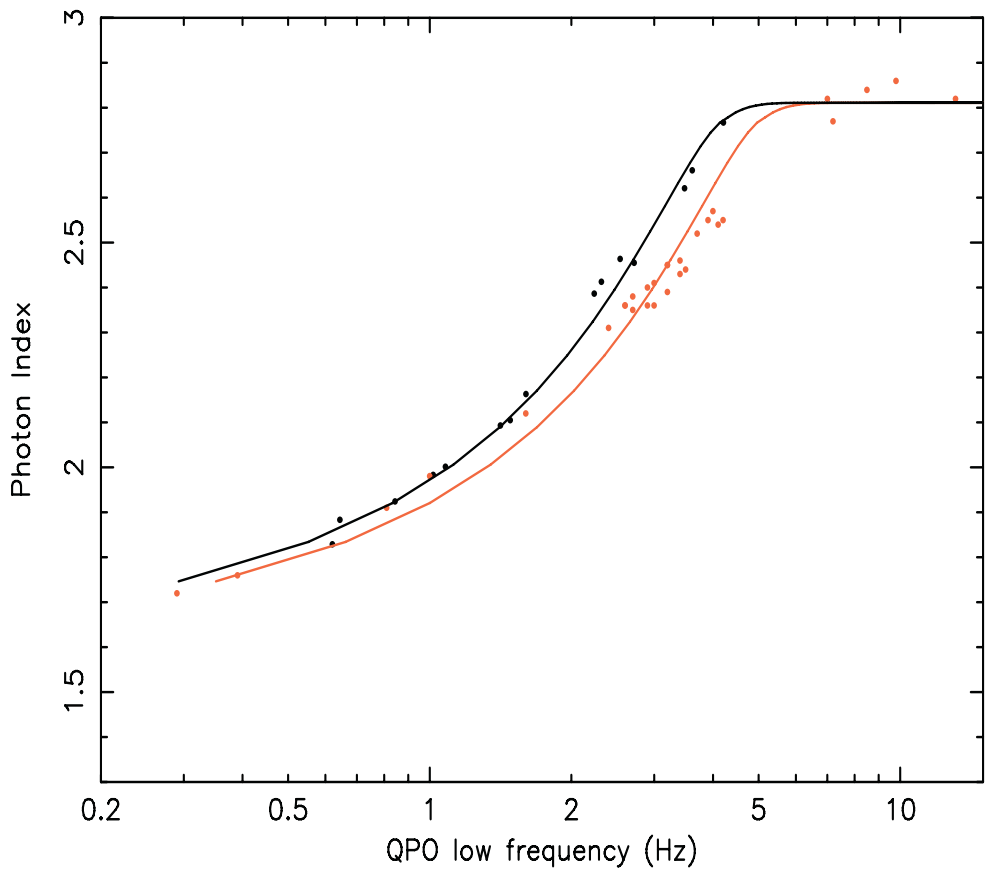}
\caption{
Comparison of the observed (points) and the  theoretical correlations
(solid lines) of photon index vs QPO low frequency between GRS 1915+105 (observations by Vignarca et al. 2003) and
XTE J1550-564 [observations by Sobczak et al. (1999), (2000); Remillard et al. (2002a,b), see also 
  Fig. 6, 8 in Vignarca et al.]. Black points and line for GRS 1915+105 and red points and line
for XTE J1550-564. The XTE J1550-564 curve is produced by sliding the GRS
1915+105 curve along the frequency axis with factor $12/10$ (see text for
details). }
\label{1915_1550}
\end{figure}
\newpage
\begin{figure}[ptbptbptb]
\includegraphics[width=4.5in,height=5.1in,angle=-90]{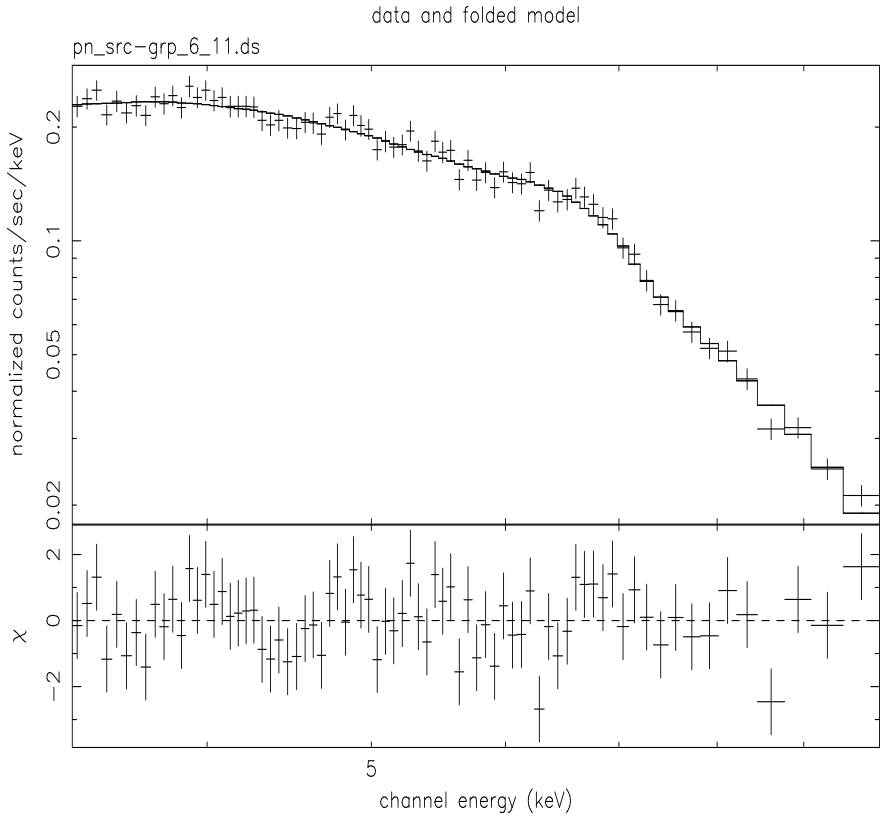}
\caption{
BMC fit to the XMM PN energy spectrum of M82-X1 in the range 3.3 to
10 keV.
}
\label{M82-X1}
\end{figure}
\newpage
\begin{figure}[ptbptbptb]
\includegraphics[width=3.4 in,height=3.3in,angle=90]{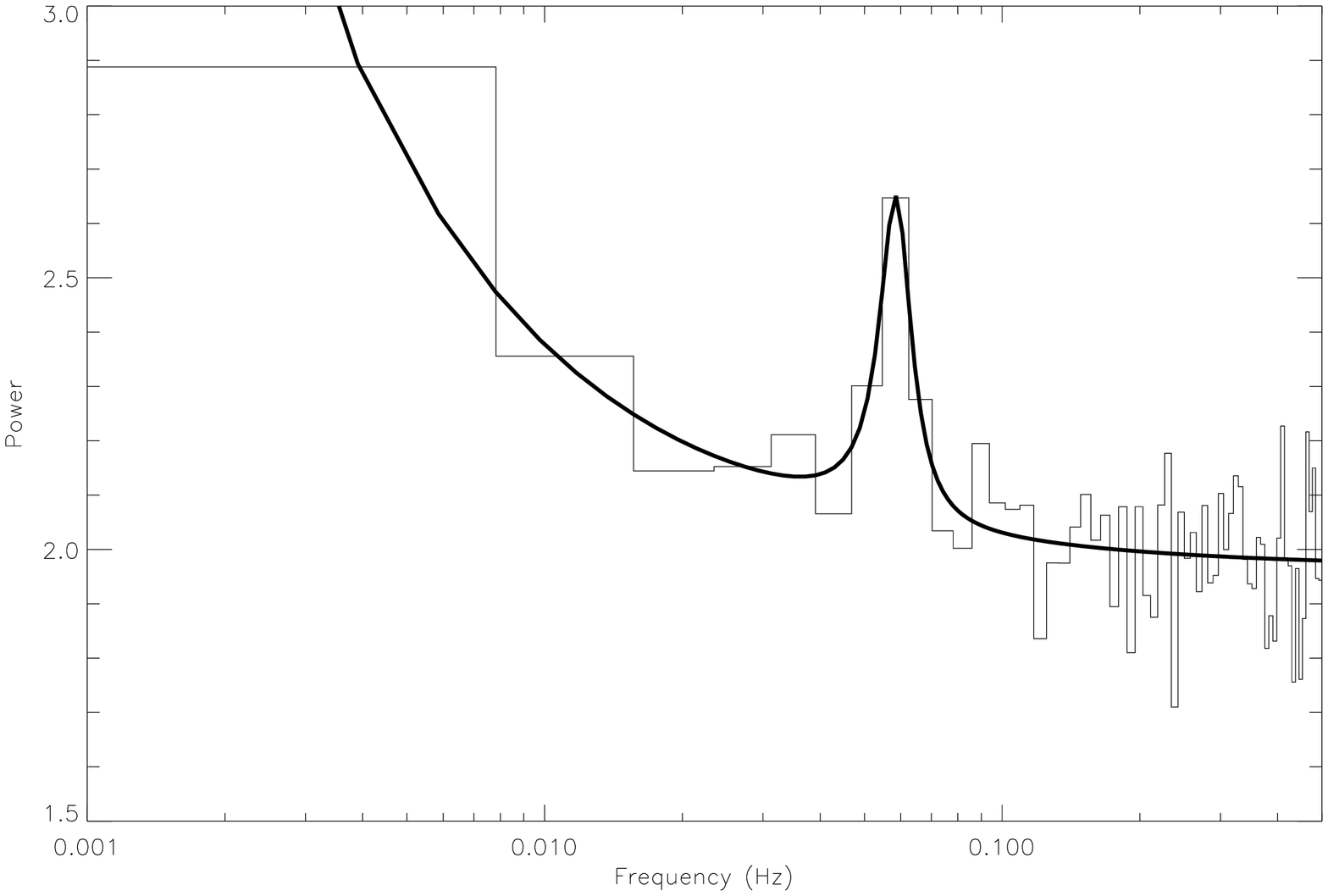}
\includegraphics[width=2.6in,height=3.in,angle=90]{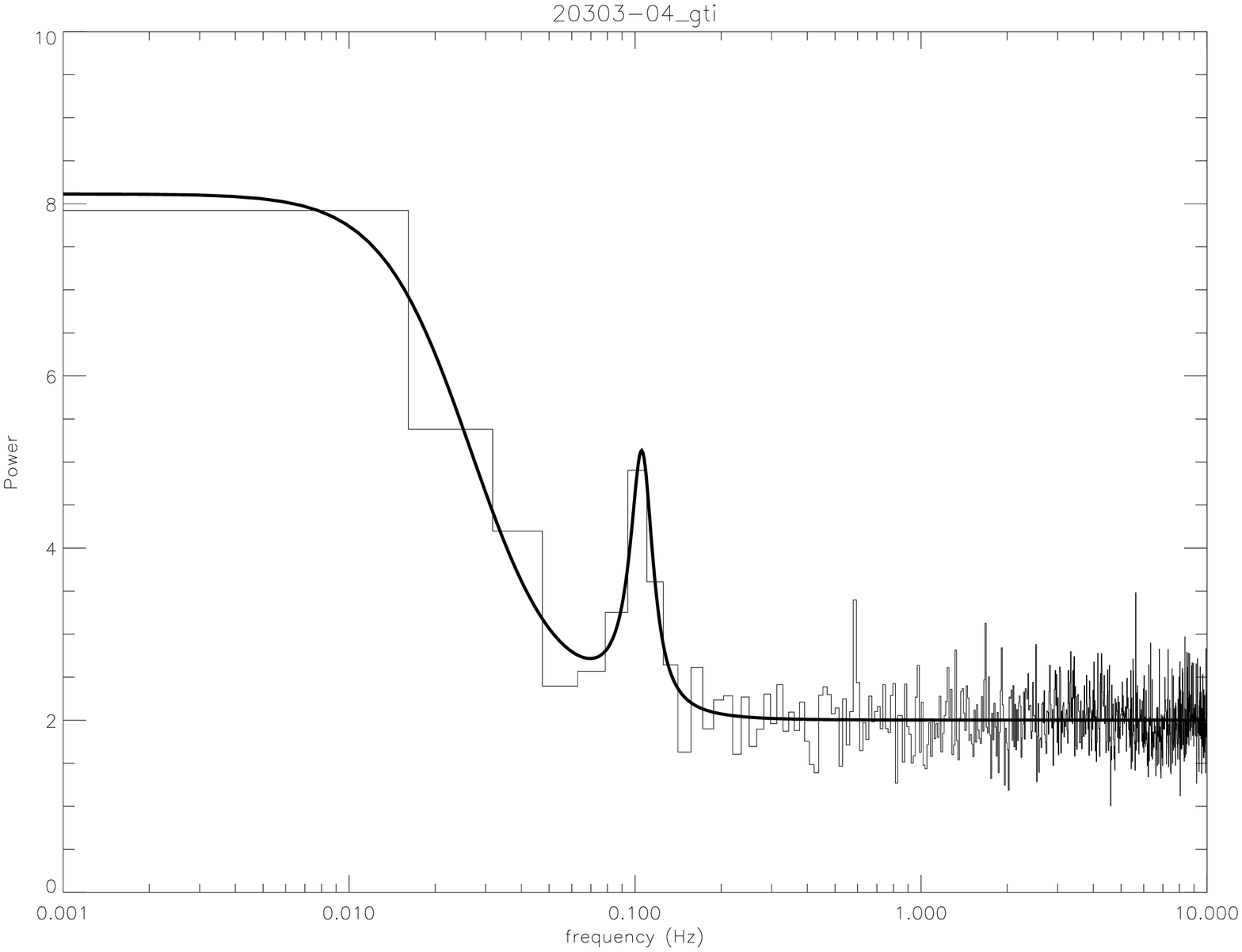}
\caption{
Left panel: Power spectrum from XMM EPIC PN instrument for the source 
M82-X1  for the  energy range 3.3-10keV showing  a QPO at 57 mHz.
Right panel:Same for RXTE OBSID 20303-02-04 showing a break
frequency at 26 mHz and QPO frequency at 106 mHz. 
}
\label{PDS}
\end{figure}

\end{document}